\documentclass[hessletter]{aa}

\usepackage{graphicx}
\usepackage{txfonts}
\usepackage{color}
\usepackage{natbib}
\usepackage{tabularx}
\usepackage{multirow}
\bibpunct{(}{)}{;}{a}{}{,} % to follow the A&A style

\begin{document}

\title{Very-high-energy gamma-ray emission from the direction of the 
Galactic globular cluster Terzan~5}

\author{HESS Collaboration
\and A.~Abramowski \inst{1}
\and F.~Acero \inst{2}
\and F.~Aharonian \inst{3,4,5}
\and A.G.~Akhperjanian \inst{6,5}
\and G.~Anton \inst{7}
\and A.~Balzer \inst{7}
\and A.~Barnacka \inst{8,9}
\and U.~Barres~de~Almeida \inst{10}\thanks{supported by CAPES Foundation, Ministry of Education of Brazil}
\and Y.~Becherini \inst{11,12}
\and J.~Becker \inst{13}
% ...
}

\institute{home}

\institute{\tiny 
Universit\"at Hamburg, Institut f\"ur Experimentalphysik, Luruper Chaussee 149, D 22761 Hamburg, Germany \and
Laboratoire Univers et Particules de Montpellier, Universit\'e Montpellier 2, CNRS/IN2P3,  CC 72, Place Eug\`ene Bataillon, F-34095 Montpellier Cedex 5, France \and
Max-Planck-Institut f\"ur Kernphysik, P.O. Box 103980, D 69029 Heidelberg, Germany \and
Dublin Institute for Advanced Studies, 31 Fitzwilliam Place, Dublin 2, Ireland \and
National Academy of Sciences of the Republic of Armenia, Yerevan  \and
Yerevan Physics Institute, 2 Alikhanian Brothers St., 375036 Yerevan, Armenia \and
Universit\"at Erlangen-N\"urnberg, Physikalisches Institut, Erwin-Rommel-Str. 1, D 91058 Erlangen, Germany \and
Nicolaus Copernicus Astronomical Center, ul. Bartycka 18, 00-716 Warsaw, Poland \and
CEA Saclay, DSM/IRFU, F-91191 Gif-Sur-Yvette Cedex, France \and
University of Durham, Department of Physics, South Road, Durham DH1 3LE, U.K. \and
Astroparticule et Cosmologie (APC), CNRS, Universit\'{e} Paris 7 Denis Diderot, 10, rue Alice Domon et L\'{e}onie Duquet, F-75205 Paris Cedex 13, France \thanks{(UMR 7164: CNRS, Universit\'e Paris VII, CEA, Observatoire de Paris)} \and
Laboratoire Leprince-Ringuet, Ecole Polytechnique, CNRS/IN2P3, F-91128 Palaiseau, France \and
Institut f\"ur Theoretische Physik, Lehrstuhl IV: Weltraum und Astrophysik, Ruhr-Universit\"at Bochum, D 44780 Bochum, Germany \and
Landessternwarte, Universit\"at Heidelberg, K\"onigstuhl, D 69117 Heidelberg, Germany \and
Institut f\"ur Physik, Humboldt-Universit\"at zu Berlin, Newtonstr. 15, D 12489 Berlin, Germany \and
LUTH, Observatoire de Paris, CNRS, Universit\'e Paris Diderot, 5 Place Jules Janssen, 92190 Meudon, France \and
LPNHE, Universit\'e Pierre et Marie Curie Paris 6, Universit\'e Denis Diderot Paris 7, CNRS/IN2P3, 4 Place Jussieu, F-75252, Paris Cedex 5, France \and
Institut f\"ur Astronomie und Astrophysik, Universit\"at T\"ubingen, Sand 1, D 72076 T\"ubingen, Germany \and
Astronomical Observatory, The University of Warsaw, Al. Ujazdowskie 4, 00-478 Warsaw, Poland \and
Unit for Space Physics, North-West University, Potchefstroom 2520, South Africa \and
Laboratoire d'Annecy-le-Vieux de Physique des Particules, Universit\'{e} de Savoie, CNRS/IN2P3, F-74941 Annecy-le-Vieux, France \and
Oskar Klein Centre, Department of Physics, Stockholm University, Albanova University Center, SE-10691 Stockholm, Sweden \and
University of Namibia, Department of Physics, Private Bag 13301, Windhoek, Namibia \and
Laboratoire d'Astrophysique de Grenoble, INSU/CNRS, Universit\'e Joseph Fourier, BP 53, F-38041 Grenoble Cedex 9, France  \and
Department of Physics and Astronomy, The University of Leicester, University Road, Leicester, LE1 7RH, United Kingdom \and
Instytut Fizyki J\c{a}drowej PAN, ul. Radzikowskiego 152, 31-342 Krak{\'o}w, Poland \and
Institut f\"ur Astro- und Teilchenphysik, Leopold-Franzens-Universit\"at Innsbruck, A-6020 Innsbruck, Austria \and
Obserwatorium Astronomiczne, Uniwersytet Jagiello{\'n}ski, ul. Orla 171, 30-244 Krak{\'o}w, Poland \and
Toru{\'n} Centre for Astronomy, Nicolaus Copernicus University, ul. Gagarina 11, 87-100 Toru{\'n}, Poland \and
School of Chemistry \& Physics, University of Adelaide, Adelaide 5005, Australia \and
Charles University, Faculty of Mathematics and Physics, Institute of Particle and Nuclear Physics, V Hole\v{s}ovi\v{c}k\'{a}ch 2, 180 00 Prague 8, Czech Republic \and
School of Physics \& Astronomy, University of Leeds, Leeds LS2 9JT, UK \and
European Associated Laboratory for Gamma-Ray Astronomy, jointly supported by CNRS and MPG \and
Oskar Klein Centre, Department of Physics, Royal Institute of Technology (KTH), Albanova, SE-10691 Stockholm, Sweden}

\offprints{\email{\\clapson@mpi-hd.mpg.de,domainko@mpi-hd.mpg.de}} 

\date{Received May 2, 2011; accepted June 16, 2011}
 
\abstract{The H.E.S.S. very-high-energy (VHE, E $>$ 0.1 TeV) gamma-ray telescope
system has discovered a new source, \object{HESS~J1747-248}. The measured integral flux
is $(1.2\pm0.3) \times 10^{-12}$ cm$^{-2}$s$^{-1}$ above 440 GeV for a
power-law photon spectral index of $2.5\pm0.3_\mathrm{stat}\pm0.2_\mathrm{sys}$. 
The VHE gamma-ray source is located in the close vicinity of the Galactic
globular cluster \object{Terzan~5} and extends beyond the H.E.S.S. point spread function (0.07\degr).
The probability of a chance coincidence with Terzan~5 and an unrelated VHE
source is quite low ($\sim 10^{-4}$). With the largest population of
identified millisecond pulsars (msPSRs), a very high core stellar density
and the brightest GeV range flux as measured by \emph{Fermi}-LAT, Terzan~5
stands out among Galactic globular clusters. The properties of the VHE
source are briefly discussed in the context of potential emission
mechanisms, notably in relation to msPSRs. Interpretation of the available
data accommodates several possible origins for this VHE gamma-ray source,
although none of them offers a satisfying explanation of its peculiar morphology.}


\maketitle

\titlerunning{VHE emission from Terzan~5}
\authorrunning{HESS Collaboration}

\section{Introduction}

Several types of Galactic VHE $\gamma$-ray sources, such as pulsar wind
nebulae (PWNe) and supernova remnants, have been detected to date, but 
so far none are in the vicinity of a globular cluster (GC).
GCs are very old stellar systems with exceptionally 
high densities of stars in their cores, leading to
numerous stellar collisions
\citep[see e.g.][]{pooley2006}, and they also contain many 
millisecond pulsars \citep[msPSRs,][]{ransom2008}, likely 
related to the large number of binary stellar members \citep{camilo2005}.
GCs are predicted to emit VHE $\gamma$-rays 
from inverse Compton (IC) up-scattering of photons from both
stellar radiation fields and the cosmic microwave background by energetic electrons. 
Electrons could be produced 
at least at two different sites: in the magnetosphere 
of the msPSRs \citep[][hereafter VJ09]{venter2009} 
or at msPSR wind nebulae shocks where electrons produced by 
the msPSRs \citep[][hereafter BS07]{bednarek2007} could be re-accelerated.
Recently observed
GeV $\gamma$-ray emission from several GCs
\citep[\emph{Fermi}-LAT observations: ][]{abdo2009,kong2010,abdo2010,tam2011}
prove that electrons accelerated up to the required energy range are present in GCs.

This paper reports the discovery of VHE
$\gamma$-rays from the direction of the GC Terzan~5. 
Terzan~5  hosts the largest population of msPSRs detected in a GC so far 
\citep[33,][]{ransom2008}. It also contains 
the largest estimated number of msPSRs as derived from its 
emission in the GeV \citep[180$^{+100}_{-90}$,][]{abdo2010} 
and radio domains \citep[60~--~200,][]{fruchter2000,kong2010}. 
It is located at a distance of 5.9~kpc \citep{ferraro2009}
at RA(J2000)~17$^\mathrm{h}$48$^\mathrm{m}$04$^\mathrm{s}$.85 and 
Dec~-24\degr46\arcmin44\arcsec.6 
(Galactic coordinates: $l = 3.84\degr$, $b = 1.69\degr$)
%Dec~$-24^{\circ}$46$^\prime$44$^{\prime\prime}$.6 
%(Galactic coordinates: $l = 3.84^{\circ}$, $b = 1.69^{\circ}$)
and exhibits a core radius $r_\mathrm{c} = 0\arcmin.15$, %0^\prime.15$, 
a half-mass radius $r_\mathrm{h} = 0\arcmin.52$, %0^\prime.52$ 
and a tidal radius $r_\mathrm{t} = 4\arcmin.6$ % 4^\prime.6$ 
\citep{lanzoni2010}. 
The detection of diffuse non-thermal X-ray emission
centred on the GC but extending beyond $r_\mathrm{h}$ (between 1 and 5.8 $r_\mathrm{h}$)
has been reported by \citet{eger2010}. 
Observations of other GCs in the VHE domain so far have only resulted in upper limits 
\citep[on \object{M~13}, \object{M~15}, \object{M~5}, and \object{47~Tucanae} in][respectively, and references therein for older results]{anderhub2009,mccutcheon2009,aharonian2009}. 

%{\color{blue}
%This paper is organized as follows: in Sec.~\ref{subsection:analysis} the 
%observation and analysis results are presented. 
%The probability of a chance coincidence is explored 
%in Sec.~\ref{subsection:chance}. 
%Scenarios for the VHE $\gamma$-ray production are discussed
%in Sec.~\ref{subsection:leptonic} for leptonic models  
%in the context of Galactic GCs
%and in Sec.~\ref{subsection:hadronic} for a hadronic model.
%}

%\input{hessdata_answer3.tex}
\section{Observation and analysis}
\label{subsection:analysis}

H.E.S.S. is an array of four Imaging Atmospheric Cherenkov Telescopes,
located in the Khomas Highland of Namibia (latitude 23.16\degr %$^{\circ}$
South, altitude 1800~m). 
Stereoscopic trigger and analysis methods allow efficient background 
(cosmic ray, CR) rejection and accurate reconstruction of energy (better than 20\%)
and arrival direction (better than 0.1\degr %$^\circ$ 
per event) for $\gamma$-rays  
in the range 0.1~--~100~TeV.
For point-like sources, the array has a nominal detection sensitivity 
of $\sim$1\% of the flux of the Crab
Nebula (Crab) above 1~TeV with a significance of 5~$\sigma$ 
in 25~hours of observation at small zenith angles ($<20\degr$). %^\circ$).
A thorough discussion of the H.E.S.S. standard analysis 
based on Hillas parameters and the performance of
the instrument can be found in \citet{aharonian2006}.
Data on Terzan~5 was obtained by H.E.S.S. from 2004 to 2010 both as part of its systematic survey
of the Galactic plane and with dedicated observations.

\begin{figure}[ht]
\centering
\includegraphics[width=9cm]{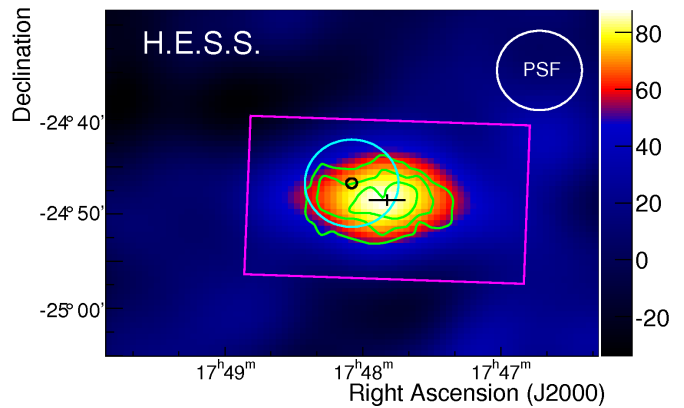}
%{T5_aug10_25deg-allweather_t34_template_releaseHAP_hard_south_1b_excesscorr2.eps}
\caption{Exposure-corrected excess image from the H.E.S.S. data,
smoothed with a Gaussian function of width 0.1\degr %$^\circ$ 
and overlaid with significance contours (4~--~6~$\sigma$) in RADec~J2000 coordinates. 
The circles show the half-mass radius (in black)
and the larger tidal radius (in cyan) of the GC.
The cross indicates the best-fit source position of HESS~J$1747-248$,
assuming a 2D Gaussian shape,
with 1~$\sigma$ uncertainty on each axis.
The rectangle represents the integration
region used for the full-source spectral analysis. 
The upper-right corner circle illustrates the instrumental PSF.
}
\label{figure:excess_hess}
\end{figure}

For source detection, the data quality cuts 
%(adapted to maximise exposure for source detection)
%(minimum 3-telescope observation, {\color{red}some weather-related fluctuations tolerated},
%pointing offset from the target position up to 2.5$^{\circ}$)
result in 90~hours of 3- and 4-telescope data
with an average zenith angle of 20.4\degr %$^{\circ}$ 
and a mean pointing direction offset from Terzan~5 of 0.95\degr. %$^{\circ}$.
After applying \emph{hard} cuts \citep[see][]{aharonian2006} optimized for point-like sources,
with a corresponding energy threshold of 380~GeV and 
a point spread function (PSF) 68\% containment radius 
$\mathrm{r}_\mathrm{PSF} = 0.07\degr$ ($4\arcmin.2$), %0.07^{\circ}$ ($4^\prime .2$), 
a source of VHE $\gamma$-rays is detected 
\citep[see Fig.~\ref{figure:excess_hess}, produced
using the \emph{template background estimation method}, 
described in][]{rowell2003}. 
The significance reaches 5.3~$\sigma$ at the position of Terzan~5
with a nearby peak significance of 7.5~$\sigma$ pre-trial 
(above 5~$\sigma$ post-trial). 
%With these numbers, a new VHE $\gamma$-ray source is detected.
All results are confirmed by analysis based on fully independent calibration 
and analysis chains \citep{denaurois2009}.

The source appears to extend beyond the GC tidal radius.
A 2D Gaussian fit\footnote{Obtained with the data analysis pipeline CIAO~v4.2 sherpa~v2 \citep{freeman2001}: http://cxc.cfa.harvard.edu/sherpa/.}
results in a best-fit position ($\chi^2/\mathrm{ndf} = 0.18$) 
RA(J2000)~17$^\mathrm{h}$47$^\mathrm{m}$49$^\mathrm{s} \pm 1^\mathrm{m}.8_{stat} \pm 1^\mathrm{s}.3_{sys}$ and 
Dec~$-24\degr48\arcmin30\arcsec \pm 36\arcsec_\mathrm{stat}  \pm 20\arcsec_\mathrm{sys}$ 
%Dec~$-24^{\circ}$48$^\prime$30$^{\prime\prime} \pm 36^{\prime\prime}_\mathrm{stat}  \pm 20^{\prime\prime}_\mathrm{sys}$ 
\citep[see][]{acero2010a},
offset by $4\arcmin.0 \pm 1\arcmin.9$ %$4^\prime.0 \pm 2^\prime.0$ 
from the GC centre,
therefore, this new VHE $\gamma$-ray source is named HESS~J$1747-248$. 
The size of the source is given by
the Gaussian widths $9\arcmin.6 \pm 2\arcmin.4$ %9$^\prime$.6$ \pm $2$^\prime$.4
and $1\arcmin.8 \pm 1\arcmin.2$ % 1$^\prime$.8$ \pm $1$^\prime$.2, 
for the major and minor axes, respectively, 
oriented $92\degr \pm 6\degr$ % 92$^\circ \pm $6$^\circ$ 
westwards from north.

%MakeSpectrum("T5_aug10_2deg_t4_BoxRegion13_releaseHAPhard_south_1b_th36.root","PL","Chi2",0.4,24,9,0,"BinW",-2.5)

\begin{figure}[hm]
\centering
\includegraphics[width=7.5cm]{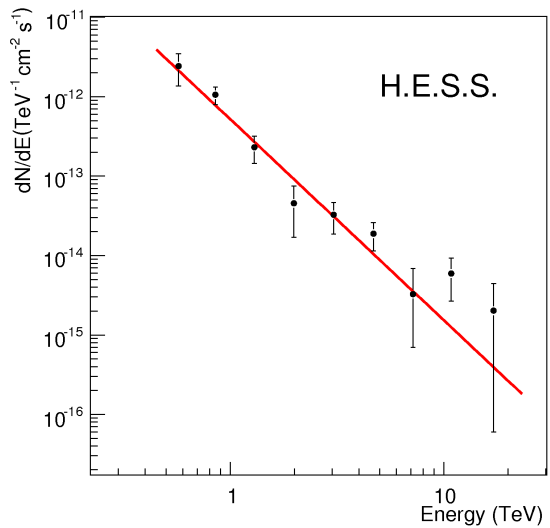}
%{Box13_hard_south_1b_th36_spec4.eps}
\caption{VHE $\gamma$-ray spectrum of HESS~J1747-248
with 1~$\sigma$ error bars, fitted with a power-law model.
The fit results are discussed in the text.
}
\label{figure:spectrum_hess}
\end{figure}

To establish the photon spectrum of the source,
a more restrictive data selection is applied,
%keeping only observations within 2$^\circ$ of the target 
to improve the energy reconstruction, 
resulting in a total of 62~h of live time.
Figure~\ref{figure:spectrum_hess}
illustrates the results of a spectral analysis
with the \emph{reflected background estimation method} \citep{berge2007}.
The test region (shown in Fig.~\ref{figure:excess_hess}) 
is a rectangle defined from the 2D Gaussian fit 
(half dimensions $r_\mathrm{PSF}+\sigma_\mathrm{x}$, $2 \times r_\mathrm{PSF}$, rotated and
centred on the best-fit position). 
Larger integration regions give lower signal significance.
For a power-law spectral model $k \left(\frac{E}{E_0} \right)^{-\Gamma}$, 
the flux normalization $k$ at 1~TeV is 
(5.2$\pm$1.1)$\times$10$^{-13}$~cm$^{-2}\,$s$^{-1}\,$TeV$^{-1}$,
and the spectral index $\Gamma = 2.5 \pm 0.3_\mathrm{stat} \pm 0.2_\mathrm{sys}$,
corresponding to an integral photon flux within the integration region of 
(1.2$\pm$0.3)$\times$10$^{-12}$~cm$^{-2}\,$s$^{-1}$, or 1.5\% of the Crab flux,
in the 0.44~--~24~TeV range ($\chi^2/\mathrm{ndf} = 1.1$). 
There are not enough excess events to discuss a more complex spectral model.

For comparison with the VHE $\gamma$-ray flux upper limit 
on the GC 47~Tucanae \citep{aharonian2009}, see Sect.~\ref{subsection:47tucan},
where a spectral analysis for a point-like source centred on the core of Terzan~5 is carried out.
It results in a compatible photon index and a reduced flux normalization,
corresponding to an integral flux in the energy range ($0.8 < \mathrm{E} < 48.6$~TeV) 
of (1.9$\pm$0.7)$\times$10$^{-13}$~cm$^{-2}\,$s$^{-1}$.
%or $\approx 0.6$\% of the Crab flux.

\section{Discussion}

In the vicinity of the discovered VHE $\gamma$-ray source,
the archival multi-wavelength (MWL) data and astronomical 
catalogues do not show any typical candidate VHE $\gamma$-ray emitter
\citep[e.g. an SNR, a young energetic pulsar or PWN, see][]{clapson2011}.
Its properties are unexpected for a GC, notably its extension, 
indication of offset ($\sim2.0 \sigma$ level) from Terzan~5
and misalignment with the centre of the GC. 
Thus the source is challenging to interpret. 

%Several approaches excluding and including the GC as the origin of the VHE
Several approaches to constrain possible origins of the VHE
radiation are considered here. First, chance coincidence probabilities,
notably with young unrelated PWNe, are derived in
Sect.~\ref{subsection:chance}.
Next, the possible emission mechanisms for GCs are addressed.
VHE leptons are covered in Sect.~\ref{subsection:leptonic}
in relation to msPSRs.
A comparison with other GCs, identified as potential VHE $\gamma$-ray sources,
is performed in Sect.~\ref{subsection:47tucan}.
Finally, possibilities for a hadronic emission scenario are discussed
in Sect.~\ref{subsection:hadronic}.

\subsection{Chance coincidence}
\label{subsection:chance}

The proximity on the sky of the GC and the VHE $\gamma$-ray source  
could simply be the chance coincidence of two physically unrelated objects. 
In particular, the source parameters (extension, photon spectrum) 
are compatible with those of identified 
VHE PWNe \citep[see e.g.][]{mattana2009}.
The probabilities of a chance coincidence 
based on the distribution of known objects are considered
in two ways.

The probability of source confusion in the VHE domain can be derived
from the distribution of source counts as a function of Galactic latitude.
\citet{chaves2009} find that this distribution, 
in the longitude range $-85\degr < l < 60\degr$, %$-85^{\circ} < l < 60^{\circ}$,
can be described by a Gaussian profile
(containing 48 sources) centred at $b=-0.26\degr$ %^{\circ}$ 
with a width $\sigma = 0.40\degr$ %^{\circ}$ 
and additionally four outliers at latitude $b < -2\degr$, %^\circ$, 
below the Galactic plane. 
There are no other detected VHE $\gamma$-ray sources 
in the latitude band $1.5\degr < b < 2\degr$. %$1.5^{\circ} < b < 2^{\circ}$.
At a latitude of 1.7\degr, %$^{\circ}$, 
Terzan~5 is almost 5~$\sigma$ away from the centre of the Gaussian,
in a region where the total number of expected sources
is negligible ($\lesssim 10^{-3}$); i.e. the source presented 
here would be an extreme outlier in the Gaussian distribution.
Using one source as the total expectation for this latitude band
over the longitude range observed by the H.E.S.S. Galactic Plane Survey, 
the probability of finding 
a VHE $\gamma$-ray source within a conservative distance of 0.1\degr %$^\circ$ 
from Terzan~5 is estimated to be $\sim 10^{-4}$.
A caveat to this argument is that the sky coverage by H.E.S.S. 
is highly non-uniform, 
with longer exposure (favouring source detection) in regions
typically located very close to the Galactic plane. 

A connection has been suggested between VHE $\gamma$-ray sources
and the PWN of pulsars 
with distance-scaled spin-down luminosity above 10$^{34}$~erg$\,$s$^{-1}\,$kpc$^{-2}$ 
\citep{carrigan2008}. They could be observable during certain
stages of their evolution with characteristics similar to the detected
VHE source \citep{mattana2009}.
In the band defined above, 
there is only one such pulsar (PSR~J1124$-$5916) in the ATNF 
catalogue\footnote{http://www.atnf.csiro.au/research/pulsar/psrcat/ (version 1.40)} 
\citep{manchester2005}.
It has been proposed that the number of radio-loud 
and radio-quiet pulsars might be comparable \citep[e.g.][]{gonthier2004},
therefore, one may expect a single undetected radio-quiet powerful pulsar.
The resulting probability of chance coincidence 
with the VHE $\gamma$-ray source
within the rectangular region shown in Fig.~\ref{figure:excess_hess} 
is again $\sim 10^{-4}$. 
This approach is limited by the strength of the association
(not every powerful pulsar is associated with a VHE $\gamma$-ray source)
and by the biases and incompleteness of the catalogues \citep[e.g.][]{gonthier2004},
illustrated by the discovery of radio-quiet GeV $\gamma$-ray pulsars 
\citep[][but none so far in the neighbourhood of Terzan~5]{abdo2010b}.

In summary, it is unlikely that the association 
of the VHE $\gamma$-ray source and the GC is by chance,
but due to simplifications in the probability estimation, 
such a possibility cannot firmly be excluded.

\subsection{Leptonic VHE $\gamma$-ray production}
\label{subsection:leptonic}

Owing to the highly energetic electrons produced by 
their msPSR populations, GCs could be $\gamma$-ray emitters,
as predicted by three different models.
In the GeV range, VJ09 %\citet{venter2009} 
explore the $\gamma$-ray emission by the superposition of pulsed
radiation from individual pulsars, whereas BS07 %\citet{bednarek2007}
and \citet{cheng2010} explore IC up-scattering by electrons either 
in the PWNe themselves or electrons that have been 
re-accelerated in colliding PWNe shocks. 
\citet{abdo2009} find that the model of VJ09 %\citet{venter2009} 
is in good agreement with the GeV properties of the \emph{Fermi}-LAT
source coincident with the GC 47~Tucanae,
while the model of BS07 %\citet{bednarek2007} 
cannot reproduce its spectral shape. 
More recently, \citet{cheng2010} have modelled 
the GeV spectra of 47~Tucanae and Terzan~5 with IC emission.

In the VHE range, these models rely on IC up-scattering of photons,
from the cluster as well as the Galactic stellar and cosmic microwave backgrounds, 
either collectively by
electrons leaving the msPSRs (see VJ09) %\citep{venter2009} 
or again by electrons re-accelerated in colliding shocks 
between PWNe (see BS07). %\citep{bednarek2007}.
The models of BS07 %\citet{bednarek2007} 
and VJ09 %\citet{venter2009}
predict a flux in the VHE range of $\sim 1$\% of the 
Crab flux for reasonable input parameters, similar to the source flux determined in this work. 
For Terzan~5, \citet{cheng2010} do not predict 
significant $\gamma$-ray emission at energies beyond 100~GeV.
Adopting the %\citet{venter2009} 
model of VJ09 with updated physical parameters for the GC
\citep{lanzoni2010} and a distance of 5.9~kpc,
a population of 220$\pm 50$~msPSRs
for a magnetic field of 10~$\mu$G (167$\pm 39$~msPSRs for 20~$\mu$G)
is required to explain the observed luminosity of the VHE  $\gamma$-ray 
source above 1~TeV. This agrees well
with the number of msPSRs derived 
in \citet{abdo2010} in the GeV range.
The number of required msPSRs is quite insensitive
to the magnetic field in the range 10~--~20~$\mu$G, although it
increases rapidly outside this range (e.g. by $\sim 50$\% at 5 $\mu$G, see VJ09).
%\citep[e.g. by about 50\% at 5 $\mu$G, see][]{venter2009}.
This model neglects $\gamma$-ray emission from
outside $r_\mathrm{h}$, due to the uncertainties in the diffusion coefficient
(and derived residence time) of the VHE electrons in this region,
likely to be unusual if the GC is a dwarf galaxy remnant as proposed by \citet{ferraro2009}.
The additional interactions included in BS07 %\citet{bednarek2007} 
result in an extended IC source of 2\arcmin~--~3\arcmin %2~--~3$^\prime$ 
radius, in rough agreement with the observed source size.
When scaling their predictions for Terzan~5 
(their Fig.~7, solid curve, top panel)
in the same way as done in \citet{aharonian2009}, 
for 180 msPSRs of individual spin-down power
10$^{34}$~erg$\,$s$^{-1}$,
the predicted integral flux is
F(E$>$440GeV)$\approx$1$\times$10$^{-12}$~cm$^{-2}\,$s$^{-1}$,
at the level of the H.E.S.S. measurement.

The IC emission should be accompanied 
by synchrotron emission in the X-ray band \citep{venter2008}. 
Non-thermal X-ray emission extending beyond $r_\mathrm{h}$ 
has indeed been discovered from Terzan~5 \citep{eger2010}. 
The nature of this X-ray emission, apparently centred on the GC, 
is still unknown and may be unrelated to HESS~J1747-248. 
Based on the energetics a synchrotron origin seems to be preferred \citep{eger2010}. 
This interpretation is challenged by
the spatial extension and the indication of an offset of the 
VHE $\gamma$-ray source from the GC centre.
A simple model where the X-ray and VHE $\gamma$-ray emission originates
in the same population of electrons cannot produce IC $\gamma$-ray
emission displaced from the peak of the stellar photon field
at the GC centre as observed.

To conclude, the observed flux -- but not the morphology --
of the VHE $\gamma$-ray source
is reasonably reproduced by the msPSRs scenarios.
%\textbf{-- but not the morphology --} of the VHE $\gamma$-ray source
%is reasonably reproduced by the msPSRs scenarios.
%\textbf{The observed VHE gamma-ray source is extended beyond the tidal radius
%of the GC, offset from its centre and its elongated shape is not aligned
%with the centre of the GC, which is difficult to interpret with current
%models for gamma-ray emission from GCs.}
To explain the morphology, 
more sophisticated models should be tested including particle trapping 
and non-uniform diffusion across the GC.
Since the optical to near-infrared stellar 
photon field should be up-scattered by the VHE electrons, 
Klein-Nishina suppression of the IC process should become significant 
at multiple TeV energies.
This would cause a steepening in the VHE $\gamma$-ray spectrum,
which should therefore not follow a pure power law. 
Statistics are currently too low to test this possibility.

\subsection{Comparison to other globular clusters}
\label{subsection:47tucan}

Under the assumption that the observed VHE $\gamma$-ray
source is related to Terzan~5, results from other GCs
in the same energy band would be valuable.
From the \emph{Fermi}-LAT 
results \citep{abdo2010}, \object{NGC~6388}
seems to host a number of msPSRs comparable to Terzan~5.
Its larger radiation field, even at its greater distance, 
makes it an interesting candidate for VHE $\gamma$-ray studies \citep{abramowski2011}.
However, predictions of VHE $\gamma$-ray fluxes for GCs 
strongly depend on uncertain parameters like 
the diffusion coefficient and the cluster magnetic field, as well as on the distance from the GC centre. 

Among the other GCs observed in the VHE domain, 
47~Tucanae is a very interesting object to compare with Terzan~5.
It is located at 4~kpc (closer than Terzan~5) 
and away from the Galactic plane, thus simplifying observation at all wavelengths. 
It contains the second largest population \citep[23,][]{ransom2008} 
of individually detected msPSRs among GCs (after Terzan~5 with 33),
in agreement with the \emph{Fermi}-LAT estimate of
 33$ \pm $15~msPSRs \citep{abdo2010}.
H.E.S.S. derived a flux upper limit on 47~Tucanae
\citep{aharonian2009} 
of F(E$>$800GeV)$<$6.7$\times$10$^{-13}$cm$^{-2}$s$^{-1}$,
at 99\% confidence level, for a point-like source.
The flux from a point-like source analysis centred on Terzan~5 
in the same energy range (see Sect.~\ref{subsection:analysis}) is below this limit. 
Given the uncertainties on the correcting factors, e.g. the distance to the GCs
and the number of msPSRs, and the actual source morphology,
this comparison only illustrates that additional 
observations of GCs will be required to settle the issue of VHE $\gamma$-ray emission from their core.

\subsection{Hadronic VHE $\gamma$-ray production}
\label{subsection:hadronic}

The remnants of a type Ia supernova 
resulting from the merger of two white dwarfs could produce hadronic CRs
\citep[see e.g. the VHE $\gamma$-ray detection of SN~1006 by][]{acero2010b}. 
GCs are expected to boost the rate of stellar collisions \citep{shara2002,grindlay2006}.

The energy in hadronic CRs required to explain a VHE $\gamma$-ray source 
of luminosity $L_{\gamma}$ with a hadronic scenario is
$E_\mathrm{pp} = L_{\gamma} \tau_\mathrm{pp} \eta^{-1} (1+S)$,
where $\tau_\mathrm{pp}$ is the cooling time due to inelastic p-p collisions
(linearly dependent on the density $n$ of the ambient medium)
and $\eta \approx 1/3$ is the fraction of CR energy that is converted 
into $\pi^0$ mesons.  
CRs below the energy range probed by the observed VHE photons
are accounted for by $S$.
Here, the H.E.S.S. observation threshold of 440~GeV
translates into a minimum CR energy of 5~TeV.
With no indication of molecular material at the location 
of Terzan~5 \citep{dame2001}, 
$n \approx 0.1$~cm$^{-3}$ is assumed \citep[following][]{dickey1992}. 
For a typical SNR CR spectrum below 5~TeV of spectral index 2.0, 
\citet{atoyan2006} derive $S \approx 5$, giving 
$E_\mathrm{pp} \approx 10^{51} (n/0.1 \mathrm{cm}^{-3})^{-1} (d/5.9 \, \mathrm{kpc})^2$~erg, 
which is somewhat high for a supernova. 

An alternative scenario may be provided 
by the remnant of a short gamma-ray burst (GRB) induced by the merger of two neutron 
stars, see e.g. \citet{nakar2007} for a review and \citet{grindlay2006} in the context of GCs.
In relativistic shocks, the transfer of the initial kinetic energy to CRs 
is expected to be very efficient \citep{atoyan2006},
so the kinetic energy of a short GRB 
could roughly provide
the energy in CRs needed for a hadronic interpretation 
of the H.E.S.S. source.
Such a GRB remnant would be spatially extended, owing to the 
diffusion of CRs away from the explosion site,
in relation to its age $t \approx r^2 / 2D$, 
where $r$ is the radius of the remnant and $D$ the diffusion coefficient.

Assuming CRs of energy 5~TeV, the extension of the 
present source would give $t \approx 10^3$~yr 
for $D = 10^{28}$~cm$^{2}\,$s$^{-1}$ 
\citep[in the Galactic disk, ][]{atoyan2006}.
The rate of short GRBs in the Galaxy 
should be about one event every 
$10^4(f_\mathrm{b}^{-1}/100)^{-1}$~yr, where
$f_\mathrm{b}$ is the GRB beaming factor, 
uncertain but expected to fall in the range 
1$ \ll f_\mathrm{b}^{-1} < $100 \citep{nakar2007}.
This estimate is based on 
a rate of short GRBs of 10~Gpc$^{-3}$~yr$^{-1}$ \citep{nakar2007} 
and a density of Milky Way-type galaxies in the local Universe 
of 10$^{-2}$ galaxies per Mpc$^{-3}$ \citep{cole2001}.
For $D$ below $10^{27}$~cm$^{2}\,$s$^{-1}$ in the TeV range 
in the vicinity of the GC, as suggested by \citet{crocker2010},
the age obtained from the source extension  
is roughly compatible with the rate of short GRBs in the Galaxy.
Nevertheless, the value of $D$ in this region may differ significantly from the available estimates.

A hadronic scenario could accommodate the observed pure power-law spectrum 
in the VHE $\gamma$-ray domain, as opposed to a leptonic scenario.
Detection of thermal X-rays would strengthen the case for 
a GRB remnant, where shocks driven by sub-relativistic ejecta expelled 
during the merger would heat the interstellar 
medium \citep{domainko2005,domainko2008}.
However, no observational evidence was found in support of
either the supernova or GRB remnant scenario.

\section{Outlook}

The nature of HESS~J1747-248 is uncertain, since no counterpart or model can fully
explain the observed morphology.
An association with Terzan~5 is tantalizing, but
the available data do not firmly prove this scenario.

Several tests could be done to clarify the nature of the VHE source.
In the X-ray range, a large FoV and an extended energy range would be desirable. 
\emph{XMM-Newton} with its large FoV is suitable for searching for counterparts
over the whole extent of HESS~J1747-248. 
As a complement, the upcoming \emph{NuSTAR} mission \citep{harrison2010}
will provide significantly improved sensitivity in the $\sim10 - 70$ keV range, 
which is less affected by neutral hydrogen absorption. This may help in constraining 
the nature of the diffuse X-ray emission found by \emph{Chandra}.
Pulsar searches, with \emph{Fermi}-LAT and radio telescopes, may reveal 
a powerful pulsar in the vicinity of HESS~J1747-248. 
Finally H.E.S.S phase II, which will extend observations in the VHE $\gamma$-ray 
range to lower energies, will further characterize the source.\\
In parallel, more sophisticated source models
might narrow down the list of applicable emission scenarios.

\acknowledgements{
The support of the Namibian authorities and of the University of Namibia
in facilitating the construction and operation of H.E.S.S. is gratefully
acknowledged, as is the support by the German Ministry for Education and
Research (BMBF), the Max Planck Society, the French Ministry for Research,
the CNRS-IN2P3 and the Astroparticle Interdisciplinary Programme of the
CNRS, the U.K. Science and Technology Facilities Council (STFC),
the IPNP of the Charles University, the Polish Ministry of Science and 
Higher Education, the South African Department of
Science and Technology and National Research Foundation, and by the
University of Namibia. We appreciate the excellent work of the technical
support staff in Berlin, Durham, Hamburg, Heidelberg, Palaiseau, Paris,
Saclay, and in Namibia in the construction and operation of the
equipment.
}

\end{document}